\documentclass[conference]{IEEEtran}
\IEEEoverridecommandlockouts
\usepackage{cite}
\usepackage{amsmath,amssymb,amsfonts,subfigure}
\usepackage{algorithmic}
\usepackage{graphicx}
\usepackage{textcomp}
\usepackage{xcolor}
\def\BibTeX{{\rm B\kern-.05em{\sc i\kern-.025em b}\kern-.08em
    T\kern-.1667em\lower.7ex\hbox{E}\kern-.125emX}}

\newtheorem{theorem}{Theorem}[section]

\begin{document}

\title{Naive Bayes with Correlation Factor for Text Classification Problem
}

\author{\IEEEauthorblockN{Jiangning Chen}
\IEEEauthorblockA{\textit{School of Mathematics} \\
\textit{Georgia Institute of Technology}\\
Atlanta, US \\
jchen444@math.gatech.edu}
\and
\IEEEauthorblockN{Zhibo Dai}
\IEEEauthorblockA{\textit{School of Mathematics} \\
\textit{Georgia Institute of Technology}\\
Atlanta, US \\
zdai37@gatech.edu}
\and
\IEEEauthorblockN{Juntao Duan}
\IEEEauthorblockA{\textit{School of Mathematics} \\
\textit{Georgia Institute of Technology}\\
Atlanta, US \\
jt.duan@gatech.edu}
\and
\IEEEauthorblockN{Heinrich Matzinger}
\IEEEauthorblockA{\textit{School of Mathematics} \\
\textit{Georgia Institute of Technology}\\
Atlanta, US \\
matzi@math.gatech.edu}
\and
\IEEEauthorblockN{Ionel Popescu}
\IEEEauthorblockA{\textit{School of Mathematics} \\
\textit{Georgia Institute of Technology}\\
Atlanta, US \\
ipopescu@math.gatech.edu}

}

\maketitle

\begin{abstract}
Naive Bayes estimator is widely used in text classification problems. However, it doesn't perform well with small-size training dataset. We propose a new method based on Naive Bayes estimator to solve this problem. A correlation factor is introduced to incorporate the correlation among different classes. Experimental results show that our estimator achieves a better accuracy compared with traditional Naive Bayes in real world data.
\end{abstract}

\begin{IEEEkeywords}
Naive Bayes, correlation factor, text classification, insufficient training set
\end{IEEEkeywords}

\section{Introduction}

Text classification problem has long been an interesting research field, the aim of text classification is to develop algorithm to find the categories of given documents. Text classification has many applications in natural language processing (NLP), such as spam filtering, email routing, and sentimental analysis. Despite intensive work, there still remains  an  open  problem  today. 

This problem has been studied from many aspects, including: supervised classification problem, if we are given the labeled training data; unsupervised clustering problem, if we only have documents without labeling; feature selection.

For supervised problem, if we assume that all the categories follow independent multinomial distributions, and each document is a sample generated by that distribution. Then a straight forward idea is to use some linear models to distinguish them, such as support vector machine (SVM)\cite{cortes1995support,joachims1998text}, which is used to find the "maximum-margin hyper-plane" that divides the documents with different labels. The algorithm is defined so that the distance between the hyper-plane and the nearest sample $d_i$ from each group is maximized. The hyper-plane can be written as the set of documents vector $\vec{d}$ satisfying:
\begin{equation*}
    \vec{w}\cdot\vec{d} - b = 0,
\end{equation*}
where $\vec{w}$ is the normal vector to the hyper-plane. Under the same assumption, another effective classifier, using scores based on the probability of given documents conditioned on categories, is called Naive Bayesian classifier\cite{friedman1997bayesian,langley1992analysis, chen2018centroid}. This classifier learns from training data to estimate the distribution of each categories, then we can compute the conditional probability of each document $d_i$ given the class label $C_i$ by applying Bayes rule, then the prediction of the class is done by choosing the highest posterior probability. The algorithm to get the label for a given document $d$ is given by:
\begin{equation*}
    label(d) = \operatorname*{argmax}_j P(C_j)P(d|C_j).
\end{equation*}
Given a huge data set, we also consider using deep learning models such as Recurrent Neural Network (RNN)\cite{tang2015document,liu2016recurrent} to do classification, which includes more information such as the order of words and semantic representations.

For unsupervised problem, we have traditional method SVD (Singular Value Decomposition)\cite{albright2004taming} for the dimension reduction and clustering. There also exist some algorithms based on EM algorithm, such as pLSA (Probabilistic latent semantic analysis)\cite{hofmann1999probabilistic}, which considers the probability of each co-occurrence as a mixture of conditionally independent multinomial distributions:
\begin{eqnarray*}
    P(w,d) &=& \sum_C P(C) P(d|C) P(w|C) \\
    &=& P(d) \sum_C P(C|d) P(w|C),
\end{eqnarray*}
where $w$ and $d$ are observed words and documents, and $C$ is the words' topic. As mentioned above, parameters here are learned by EM algorithm. Using the same idea, but assuming that the topic distribution has sparse Dirichlet prior, we have algorithm LDA (Latent Dirichlet allocation)\cite{blei2003latent}. The sparse Dirichlet priors encode the intuition that documents cover only a small set of topics and that topics use only a small set of words frequently. In practice, this results in a better disambiguation of words and a more precise assignment of documents to topics.

Naive Bayes estimator is a widely used estimator, however, it requires plenty of well labeled data for training purposes. To tackle this problem, this paper proposes a novel estimation method. In the remainder of this paper, we firstly summarize the Naive Bayes estimator in section \ref{sect:Naive_Bayes}. Then we discuss the error of the Naive Bayes estimator in Theorem \ref{nb_property} and demonstrate that it is unbiased. In section \ref{sect:correlation_Naive_Bayes}, we propose a novel estimation method (see equation \ref{our_estimator}) called Naive Bayes with correlation factor. It addresses the problem in many real world text classification applications that have only limited available training data.  Furthermore, in theorem \ref{theorem:our estimator}  we show the error of the new estimator is controlled by the correlation factor and the variation has a smaller order compared with Naive Bayes estimator. In section \ref{experiment}, we show results of simulations, which demonstrates the performance of our method presented in section \ref{sect:correlation_Naive_Bayes}. Finally section \ref{conclusion} concludes our work and mentions possible future work.

\section{General Setting}
Consider a classification problem with the sample (document) set $S$, and the class set $C$ with $k$ different classes: $$C = \{C_1,C_2,...,C_k\}.$$ Assume we have totally $v$ different words, thus for each document $d\in S$, we have:
$$d = \{x_1,x_2,\cdots,x_v\}.$$
Define $y = (y_1,y_2,\cdots,y_k)$ as our label vector. For document $d$ is in class $C_i$, we have $y_i(d)=1$. Notice that for a single label problem, we have: $\sum_{i = 1}^k y_i = 1$.\\
For a test document $d$, our target is to predict: $$\hat{y}(d) = f(d;\theta) = (f_1(d;\theta),f_2(d;\theta),...,f_k(d;\theta))$$ given training sample set $S$, where $\theta$ is the parameter matrix and $f_i(d;\theta)$ is the likelihood function of document $d$ in class $C_i$. \\

\section{Naive Bayes classifier in text classification problem} \label{sect:Naive_Bayes}
In this section we will discuss the properties of estimator derived from traditional Naive Bayes method. Let class $C_i \;(1\le i\le k)$ with centroid $\theta_i = (\theta_{i_1},\theta_{i_2},...,\theta_{i_v})$ and $\theta_i$ satisfies: $\sum_{j=1}^v \theta_{i_j} = 1$. Assuming independence of the words, the most likely class for a document $d$ is computed as:
\begin{eqnarray}
label(d) &=& \operatorname*{argmax}_i P(C_i)P(d|C_i)\\
         &=& \operatorname*{argmax}_i P(C_i)\prod_{j=1}^{v} (\theta_{i_j})^{x_j}\nonumber\\
         &=& \operatorname*{argmax}_i \log{P(C_i)} + \sum_{j=1}^{v} x_j\log{\theta_{i_j}}\nonumber.
\end{eqnarray}
This gives the classification criteria once $\theta$ is estimated, namely finding the largest among
\begin{equation*}
    \log f_i(d;\theta) = \log{P(C_i)} + \sum_{j=1}^{v} x_j\log{\theta_{i_j}} \; \quad 1\le i \le k
\end{equation*}
Now we shall derive an maximum likelihood estimator for $\theta$. For a class $C_i$, we have the standard likelihood function: 
\begin{eqnarray}\label{nb_likelihood}
L(C_i,\theta) &=& \prod_{d\in S} f_i(d;\theta)^{y_i(d)}\nonumber\\
&=& \prod_{d\in C_i}\prod_{j=1}^v \theta_{i_j}^{x_j}
\end{eqnarray}
Take logarithm for both sides, we obtain the log-likelihood function:
\begin{equation}\label{nb_log_likelihood}
\log{L(C_i,\theta)} = \sum_{d\in C_i}\sum_{j=1}^v x_j\log{\theta_{i_j}}.
\end{equation}
We would like to solve optimization problem:
\begin{eqnarray}\label{nb_optimal_prob}
\max\ & &\log L(C_i,\theta)\\
\text{subject to}: & &\sum_{j = 1}^v \theta_{i_j} = 1 \nonumber \\
& & \theta_{i_j}\geq 0 \nonumber
\end{eqnarray}
The problem \eqref{nb_optimal_prob} can be explicitly solved  by Lagrange Multiplier, for class $C_i$, we have $\theta_{i} = \{\theta_{i_1},\theta_{i_2},...,\theta_{i_v}\}$, where:
\begin{equation}\label{nb_estimator}
\hat{\theta}_{i_j} = \frac{\sum_{d\in C_i}x_j}{\sum_{d\in C_i}\sum_{j=1}^v x_j}.
\end{equation}
For estimator $\hat{\theta}$, we have following theorem.

\begin{theorem}\label{nb_property}
Assume we have normalized length of each document, that is: $\sum_{j=1}^v x_j = m$ for all documents $d\in S$, the estimator \eqref{nb_estimator} satisfies following properties:
\begin{enumerate}
    \item   
    $\hat{\theta}_{i_j}$ is unbiased.
    \item   
    $E[|\hat{\theta}_{i_j}-\theta_{i_j}|^2] = \frac{\theta_{i_j}(1-\theta_{i_j})}{|C_i|m}$.
\end{enumerate}
\end{theorem}

\begin{IEEEproof}
With assumption $\sum_{j=1}^v x_j = m$, we can rewrite \eqref{nb_estimator} as:
$$\hat{\theta}_{i_j} = \frac{\sum_{d\in C_i}x_j}{\sum_{d\in C_i}m}=\frac{\sum_{d\in C_i}x_j}{|C_i|m}.$$
Since $d=(x_1,x_2,...,x_v)$ is multinomial distribution in class $C_i$, we have: $E[x_j] = m \theta_{i_j}$, and $E[x_j^2] = m\theta_{i_j}(1-\theta_{i_j}+m\theta_{i_j}).$
\begin{enumerate}
    \item \label{nb_unbiased}
    \begin{align*}
         E[\hat{\theta}_{i_j}]&=E[\frac{\sum_{d\in C_i}x_j}{|C_i|m}]=\frac{\sum_{d\in C_i}E[x_j]}{|C_i|m}\\
         &=\frac{\sum_{d\in C_i} m \theta_{i_j}}{|C_i|m}=\theta_{i_j}.
    \end{align*}
   
    Thus $\hat{\theta}_{i_j}$ is unbiased.
    \item
    By \eqref{nb_unbiased}, we have:
    $$E[|\hat{\theta}_{i_j}-\theta_{i_j}|^2]=E[\hat{\theta}_{i_j}^2]-2\theta_{i_j}E[\hat{\theta}_{i_j}]+\theta_{i_j}^2=E[\hat{\theta}_{i_j}^2]-\theta_{i_j}^2.$$
    Then notice
    \begin{equation}\label{nb_split_hat_theta}
    \hat{\theta}_{i_j}^2=\frac{(\sum_{d\in C_i}x_j)^2}{|C_i|^2m^2}=\frac{\sum_{d\in C_i}x_j^2+\sum_{d\ne d' \in C_i} x_j^{d}x_j^{d'}}{|C_i|^2m^2},
    \end{equation}
    where $d=(x_1^{d},x_2^{d},...,x_v^{d})$. \\
    Since:
    \begin{align*}
        E[\frac{\sum_{d\in C_i}x_j^2}{|C_i|^2m^2}]&=\frac{|C_i|m\theta_{i_j}(1-\theta_{i_j}+m\theta_{i_j})}{|C_i|^2m^2}\\
        &=\frac{\theta_{i_j}(1-\theta_{i_j}+m\theta_{i_j})}{|C_i|m}, 
    \end{align*}
    and
    \begin{align*}
        E[\frac{\sum_{d\ne d' \in C_i} x_j^{d}x_j^{d'}}{|C_i|^2m^2}]&=\frac{|C_i|(|C_i|-1)m^2\theta_{i_j}^2}{|C_i|^2m^2}\\
        &=\frac{(|C_i|-1)\theta_{i_j}^2}{|C_i|}.
    \end{align*}
    
    Plugging them into \eqref{nb_split_hat_theta} obtains:
    $$E[\hat{\theta}_{i_j}^2] = \frac{\theta_{i_j}(1-\theta_{i_j})}{|C_i|m} + \theta_{i_j}^2,$$
    thus: $E[|\hat{\theta}_{i_j}-\theta_{i_j}|^2] = \frac{\theta_{i_j}(1-\theta_{i_j})}{|C_i|m}$.
\end{enumerate}
\end{IEEEproof}

\section{Naive Bayes with correlation factor} \label{sect:correlation_Naive_Bayes}

From Theorem.\ref{nb_property}, we can see that traditional Naive Bayes estimator $\hat{\theta}$ is an unbiased estimator with variance $O(\frac{\theta_{i_j}(1-\theta_{i_j})}{|C_i|m})$. Now we will try to find an estimator, and prove that it can perform better than traditional Naive Bayes estimator.

Our basic idea is that, even for a single labeling problem, a document $d$ usually contains words from different classes, thus it should include feature from different classes. However, our label $y$ in training set does not reflect that information since only one component of $y$ is 1. Thus, we would like to replace $y$ by $y+t$ in Naive Bayes likelihood function \ref{nb_likelihood}
with some optimized $t$ to get our new likelihood function $L_1$: 
\begin{eqnarray}\label{new_likelihood}
    L_1(C_i,\theta) & = & \prod_{d\in S} f_i(d;\theta)^{y_i(d)+t}\nonumber\\
    & = & \prod_{d\in S} (\prod_{j=1}^v \theta_{i_j}^{x_j})^{y_i(d)+t}.
\end{eqnarray}

Notice that to compute $L_1$ of a given class $C_i$ in our estimator, instead of just using documents in $C_1$ as Naive Bayes estimator, we will use every $d\in S$.

Take logarithm for both sides of \ref{new_likelihood}, we obtain the log-likelihood function:
\begin{equation}\label{updated_log_likelihood}
\log{L_1(C_i, \theta)} = \sum_{d\in S}\ \left[ (y_i(d)+t)\sum_{j=1}^v x_j\log{\theta_{i_j}} \right].
\end{equation}

Similar to Naive Bayes estimator, We would like to solve optimization problem:
\begin{eqnarray}\label{updated_optimal_prob}
\max\ & &\log{L_1(C_i, \theta)}\\
\text{subject to}: & &\sum_{j = 1}^v \theta_{i_j} = 1 \nonumber \\
& & \theta_{i_j}\geq 0 \nonumber
\end{eqnarray}
Let:
$$G_i = 1 - \sum_{j=1}^v \theta_{i_j},$$ by Lagrange multiplier, we have: 

\begin{equation*}
\left\{
\begin{aligned}
&\frac{\partial \log(L_1)}{\partial \theta_{i_j}}+\lambda_i\frac{\partial G_i}{\partial \theta_{i_j}}=0\  \forall\ 1\leq i\leq k\ \forall\ 1\leq j\leq v\\
&\sum_{j=1}^v \theta_{i_j} = 1,\ \forall\ 1\leq i\leq k
\end{aligned}
\right.
\end{equation*}
plug in, we obtain:
\begin{equation}\label{L_1_nb_solutions}
\left\{
\begin{aligned}
&\sum_{d\in S}\frac{(y_i(d)+t)x_j}{\theta_{i_j}} - \lambda_i = 0,\  \forall\ 1\leq i\leq k\ \forall\ 1\leq j\leq v\\
&\sum_{j=1}^v \theta_{i_j} = 1,\ \forall\ 1\leq i\leq k
\end{aligned}
\right.
\end{equation}
Solve \eqref{L_1_nb_solutions}, we got the solution of optimization problem \eqref{updated_optimal_prob}:

\begin{equation} \label{our_estimator}
\hat{\theta}_{i_j}^{L_1} = \frac{\sum_{d\in S}(y_i(d)+t)x_j}{\sum_{j=1}^v\sum_{d\in S}(y_i(d)+t)x_j}
= \frac{\sum_{d\in S}(y_i(d)+t)x_j}{m(|C_i|+t|S|)}
\end{equation}

For estimator $\hat{\theta}_{i_j}^{L_1}$, we have the following result:

\begin{theorem}\label{theorem:our estimator}
Assume for each class, we have prior distributions $p_1,p_2,\cdots,p_k$ with $p_i=|C_i|/|S|$, and we have normalized length for each document, that is: $\sum_{j=1}^v x_j=m$. The estimator \eqref{our_estimator} satisfies following property: \begin{enumerate}
    \item 
    $\hat{\theta}_{i_j}^{L_1}$ is biased, with: $E[|\hat{\theta}_{i_j}^{L_1}-\theta_{i_j}|] = O(t)$
    \item 
    $E[|\hat{\theta}_{i_j}^{L_1}-E[\hat{\theta}_{i_j}^{L_1}]|^2] = O(\frac{1}{m|S|}).$
\end{enumerate}
\end{theorem}

\begin{IEEEproof}
\begin{enumerate}
    \item 
    With assumption $\sum_{j=1}^v x_j=m$, we have:
    \begin{eqnarray*}
    E[\hat{\theta}_{i_j}^{L_1}] &=& \frac{\sum_{d\in S}(y_i(d)+t)E[x_j]}{m(t|S|+|C_i|)}\\
    &=&\frac{\sum_{d\in S}tE[x_j]+\sum_{x\in C_i}E[x_j]}{m(t|S|+|C_i|)}\\
    &=&\frac{t\sum_{l=1}^k |C_l| \theta_{l_j}+\theta_{i_j}|C_i|}{t|S|+|C_i|} \\
    &=&\frac{t|S|\sum_{l=1}^k p_l\theta_{l_j}+\theta_{i_j}|C_i|}{t|S|+|C_i|}
    \end{eqnarray*}
    
    Thus:
    \begin{eqnarray*}
        E[|\hat{\theta}_{i_j}^{L_1}-\theta_{i_j}|]
        &=&\frac{t|S||\sum_{l=1}^k p_l\theta_{l_j}-\theta_{i_j}|}{t|S|+|C_i|} \nonumber\\
        &=& \frac{|\sum_{l=1}^k p_l\theta_{l_j}-\theta_{i_j}|}{1+p_i/t} \\
        &=& O(t).
    \end{eqnarray*}
    
    This shows our estimator is biased. The error is controlled by $t$. When $t$ converges to 0, our estimator converges to the unbiased Naive Bayes estimator. We can also derive a lower bound for the square error:
    \begin{eqnarray*}
        E[|\hat{\theta}_{i_j}^{L_1}-\theta_{i_j}|^2]
        &\ge &  (E[|\hat{\theta}_{i_j}^{L_1}-\theta_{i_j}|])^2\\
        &=& \frac{|\sum_{l=1}^k p_l\theta_{l_j}-\theta_{i_j}|^2}{(1+p_i/t)^2}
    \end{eqnarray*}

    \item
    For variance part, since 
    \[
    \hat{\theta}_{i_j}^{L_1}= \frac{\sum_{d\in S}(y_i(d)+t) x_j }{m(|C_i|+t|S|)},
    \] we have:
    \begin{align} \label{our_variance}
    & E[|\hat{\theta}_{i_j}^{L_1}-E [\hat{\theta}_{i_j}^{L_1}]|^2]\\
    = &E\left[ \left|\frac{\sum_{d\in S}(y_i(d)+t) (x_j-E[x_j])}{m(|C_i|+t|S|)} \right|^2 \right] \nonumber\\
    =& \frac{\sum_{d\in S}(y_i(d)+t)^2 E(|x_j-E[x_j]|^2)}{m^2(|C_i|+t|S|)^2} \nonumber\\
    =&  \frac{\sum_{d\in C_i}(1+t)^2 m \theta_{i_j}(1-\theta_{i_j}) }{m^2(|C_i|+t|S|)^2} \nonumber\\
    +& \frac{\sum_{d\in C_l, l\ne i}t^2 m \theta_{l_j}(1-\theta_{l_j})}{m^2(|C_i|+t|S|)^2} \nonumber\\
    =&  \frac{|C_i|(1+2t)  \theta_{i_j}(1-\theta_{i_j}) + \sum_{l=1}^k |C_l|t^2  \theta_{l_j}(1-\theta_{l_j}) }{m(|C_i|+t|S|)^2} \nonumber\\
    =& \frac{|S|p_i(1+2t)  \theta_{i_j}(1-\theta_{i_j}) + |S|\sum_{l=1}^k p_l t^2  \theta_{l_j}(1-\theta_{l_j}) }{m(|S|p_i+t|S|)^2} \nonumber\\
    =& \frac{p_i(1+2t)  \theta_{i_j}(1-\theta_{i_j}) + \sum_{l=1}^k p_l t^2  \theta_{l_j}(1-\theta_{l_j}) }{m|S|(p_i+t)^2} \\
    =& O(\frac{1}{m|S|})\nonumber
    \end{align}
    
\end{enumerate}

\end{IEEEproof}

We can see that $E[|\hat{\theta}_{i_j}^{L_1}-E[\hat{\theta}_{i_j}^{L_1}]|^2]$ is in $O(\frac{1}{|S|})$, which means it convergent faster than standard Naive Bayes $O(\frac{1}{|C_i|})$, however, since $E[|\hat{\theta}_{i_j}^{L_1}-\theta_{i_j}|]\neq 0$, it is not an unbiased estimator.

\section{Experiment}\label{experiment}

\subsection{Simulation with Fixed Correlation Factor}
\begin{figure} \centering
\subfigure[] { \label{2_m_r}
\includegraphics[width=0.9\columnwidth]{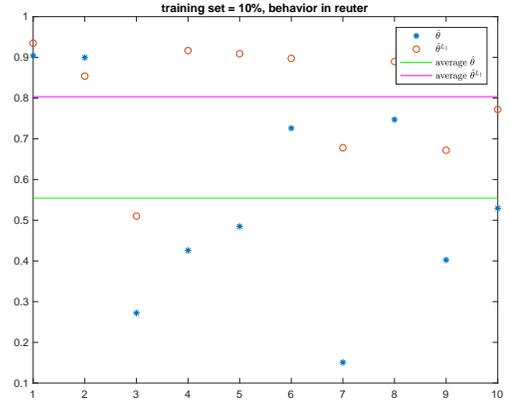}
}
\subfigure[] { \label{2_m_2}
\includegraphics[width=0.9\columnwidth]{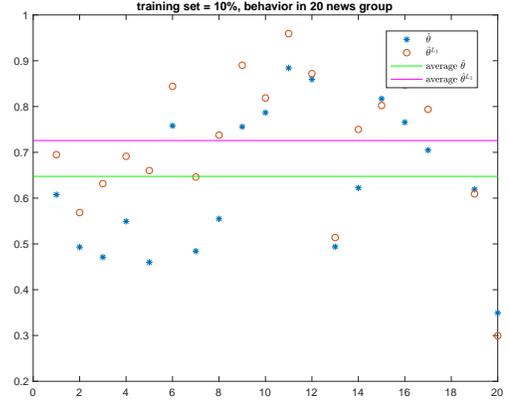} 
}
\caption{\small We take 10 largest groups in Reuter-21578 dataset (a) and 20 news group dataset (b), and take 10\% of the data as training set. The y-axis is the accuracy, and the x-axis is the class index.}
\label{2_m}
\end{figure}

\begin{figure} \centering
\subfigure[] { \label{large_r}
\includegraphics[width=0.9\columnwidth]{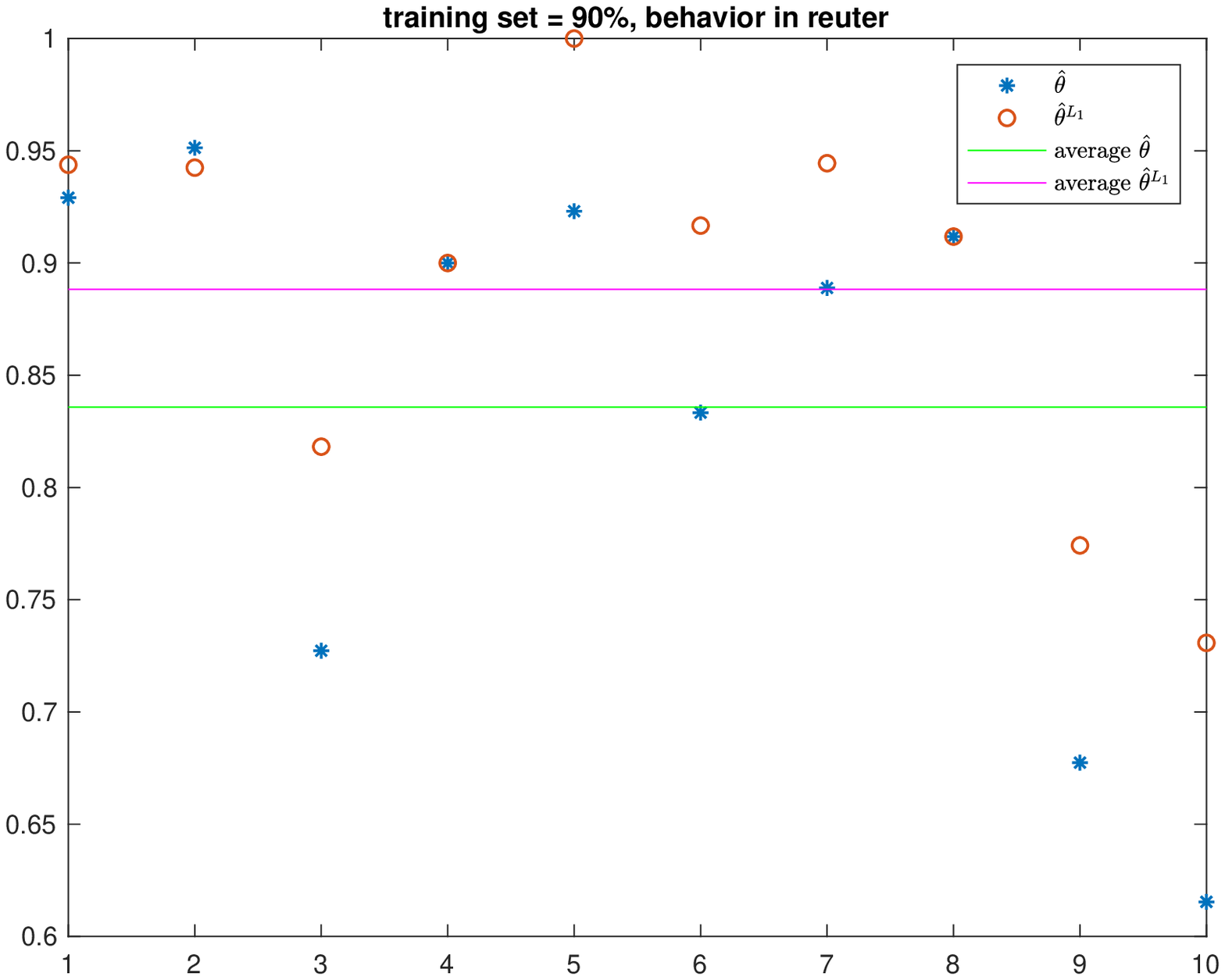}
}
\subfigure[] { \label{large_2}
\includegraphics[width=0.9\columnwidth]{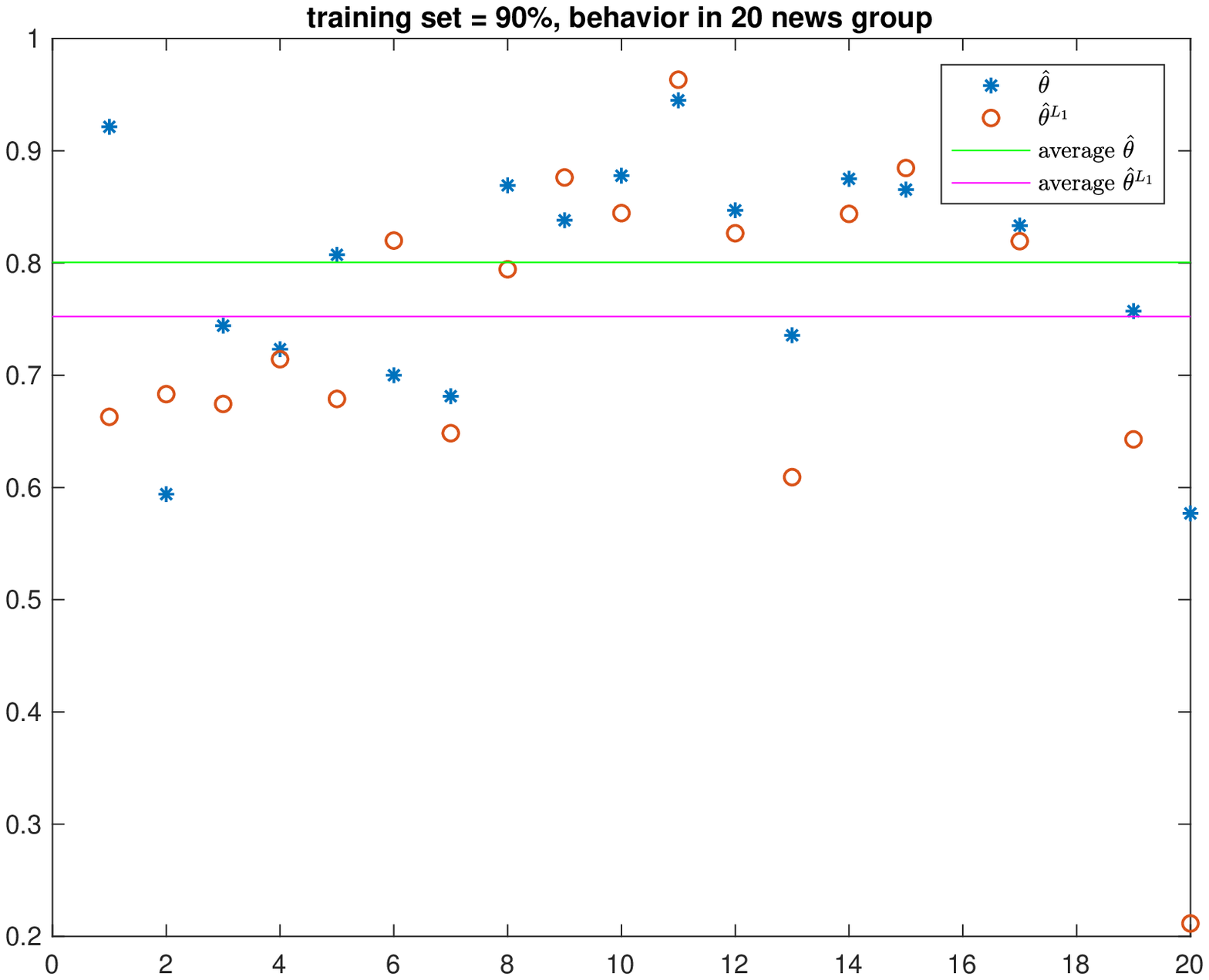} 
}
\caption{\small We take 10 largest groups in Reuter-21578 dataset (a) and 20 news group dataset (b), and take 90\% of the data as training set. The y-axis is the accuracy, and the x-axis is the class index.}
\label{large_m}
\end{figure}

\begin{figure}\centering
\subfigure[]{ \label{over_fit_r}
\includegraphics[width=0.9\columnwidth]{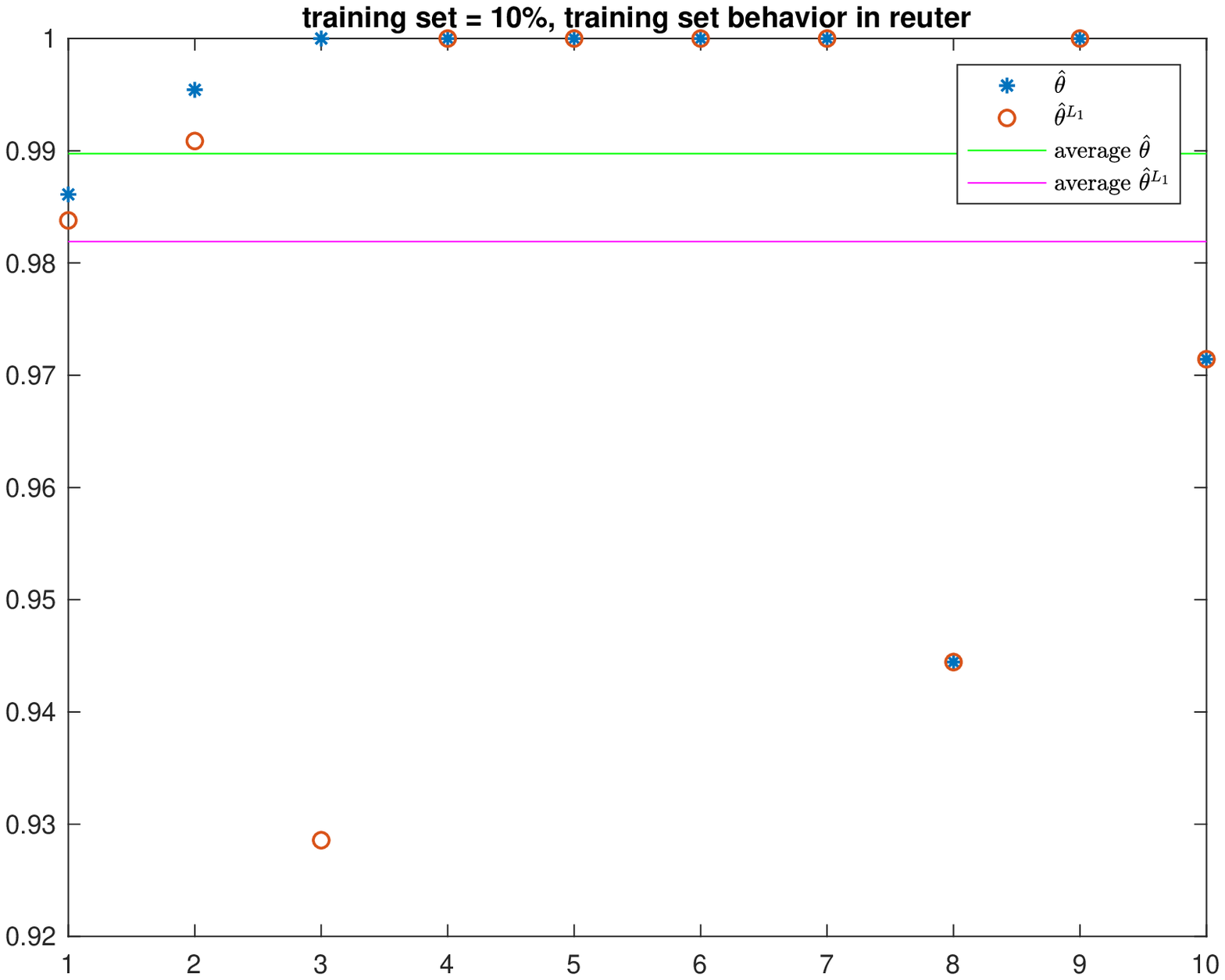}
}
\subfigure[]{ \label{over_fit_20}
\includegraphics[width=0.9\columnwidth]{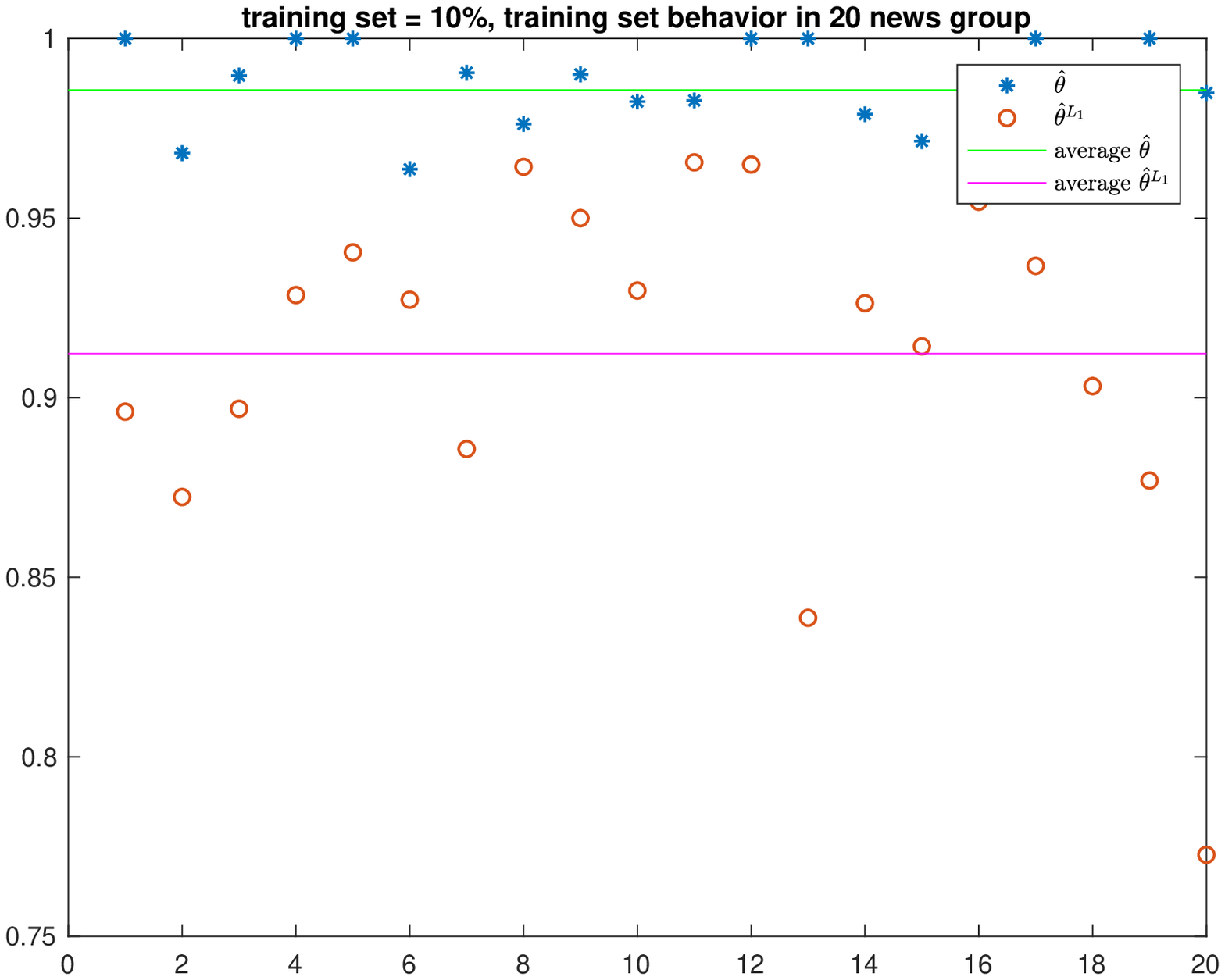}
}
\caption{We take 10 largest groups in Reuter-21578 dataset (a), and 20 news group dataset (b), and take 10\% of the data as training set. We test the result on training set. The y-axis is the accuracy, and the x-axis is the class index.}
 \label{over_fit}
\end{figure}

We applied our method on top 10 topics of single labeled documents in Reuters-21578 data\cite{reuters_data}, and 20 news group data\cite{20_news}. we compare the result of traditional Naive Bayes estimator \eqref{nb_estimator}: $\hat{\theta}_{i_j}$and our estimator \eqref{our_estimator}: $\hat{\theta}_{i_j}^{L_1}$. In this simulation,our correlation factor $t$ is chosen to be $1$ for Figure.\ref{2_m}, Figure.\ref{large_m} and Figure.\ref{over_fit}.

First of all, we run both algorithms on these two sample sets. We know that when sample size becomes large enough, our estimator actually convergences into something else. But when training set is  small, our estimator should converge faster. Thus we first take the training size relatively small. See Figure.\ref{2_m_r} and Figure.\ref{2_m_2}. According to the simulation, we can see our method is more accurate for most of the classes, and more accurate in average.

Then we test our estimator $\hat{\theta}^{L_1}$ with larger dataset. In our analysis above, we know that as dataset becomes large enough, our estimator converges to something else, so we expect a better result with traditional Naive Bayes estimator. See Figure.\ref{large_r} and Figure.\ref{large_2}. According to the simulation, we can see for 20 news group, traditional Naive Bayes performs better than our method, but our method is still more accurate than Naive Bayes in Reuter's data. The reason might be that we have a huge unbalance dataset in Reuter's data, 90\% of the training set is still not large enough for many classes.

Finally, We apply same training set with training size 10$\%$ and test the accuracy on training set instead of test set. We find traditional Naive Bayes estimator actually achieves better result, which means it might have more over-fitting problems. This might be the reason why our method works better when dataset is not too large: adding the correlation factor $t$ helps us bring some uncertainty in training process, which helps avoid over-fitting. See Figure.\ref{over_fit_r} and Figure.\ref{over_fit_20}.

\subsection{Simulation with Different Correlation Factor}
\begin{figure}\centering
\subfigure[]{ \label{diff_t_reuters}
\includegraphics[width = 0.9\columnwidth]{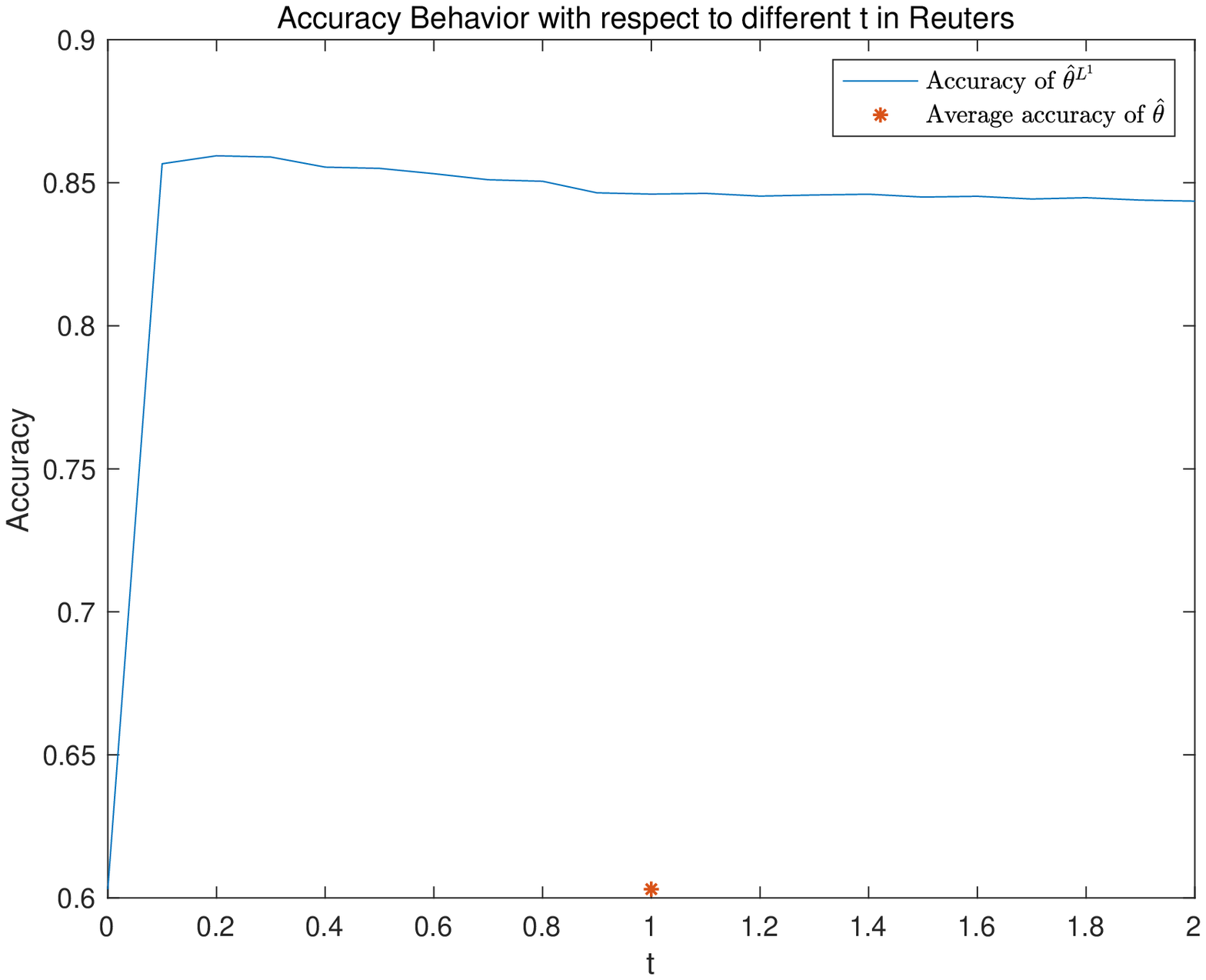}
}
\subfigure[]{ \label{diff_t_20}
\includegraphics[width = 0.9\columnwidth]{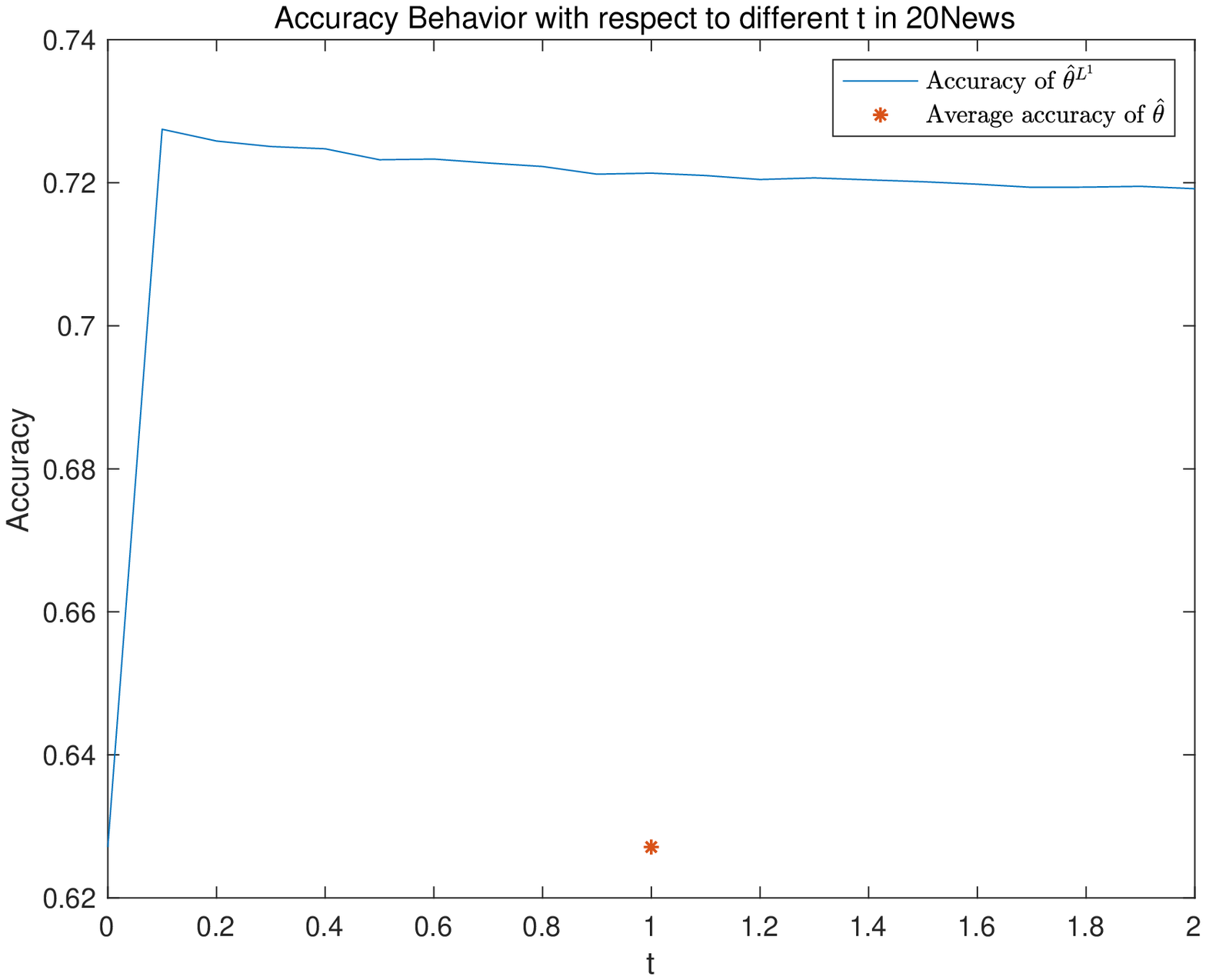}
}
\caption{We test accuracy behavior with respect to different correlation factors in Reuter-21578 (a) and 20 News group dataset (b). We take 10\% of the data as training set. The y-axis is the accuracy and the x-axis is the correlation factor $t$}
 \label{diff_t}
\end{figure}

In our estimator \eqref{our_estimator}, we need to determine how to choose correlation factor $t$. An idea is to choose $t$ to minimize the variance \eqref{our_variance}. Taking derivative of \eqref{our_variance} with respect to $t$ and setting it to be 0, we find $t$ satisfies:
\[
(p_i-t-1)\theta_{i_j}(1-\theta_{i_j}) + t\sum_{l=1}^k p_l  \theta_{l_j}(1-\theta_{l_j}) =0,
\]
that is: 
\begin{equation}\label{t_parameter}
    t=  \frac{(1-p_i)\theta_{i_j}(1-\theta_{i_j}) }{[\sum_{l=1}^k p_l  \theta_{l_j}(1-\theta_{l_j})] - \theta_{i_j}(1-\theta_{i_j})}
\end{equation}
We can see from \eqref{t_parameter} that our correlation factor $t$ should be less than 1. In our simulation, we notice that when we choose correlation factor to be around 0.1, we get best accuracy for our estimation. See Figure.\ref{diff_t_reuters} and Figure.\ref{diff_t_20}.

\section{Conclusion}\label{conclusion}
In this paper, we modified the traditional Naive Bayes estimator with a correlation factor to obtain a new estimator, which is biased but with a smaller variance. We applied our estimator in text classification problems, and showed that it works better when training data set is small.

There are several important questions related our estimator:

\begin{enumerate}
    \item We have a parameter, correlation factor $t$, in our estimator \eqref{our_estimator}. In Section \ref{experiment}, we have some simulations when $t=1$, and further show what happened when $t$ ranges from $[0,2]$, but we don't have theoretical result about how to choose $t$. One important question is how can we choose $t$ in different problems, in each of these problems, can we solve $t$ explicitly?
    \medskip
    \item We only test our result in Reuter's data \cite{reuters_data} and 20 news group \cite{20_news}, these datasets are news from newspapers, which means they are highly correlated to each other. Will our estimator still work in other more independent datasets?
    \medskip
    \item We can only use our method in single labeled dataset so far, it would be interesting to see if we can extend our result in partial labeled dataset or multi-labeled dataset.
    \medskip
\end{enumerate}

\bibliographystyle{IEEEtran}

\nocite{*}

\end{document}